\documentclass[12pt]{article}
\usepackage{graphics,cite}

\def \lsim
{\raisebox{-3pt}{$\>\stackrel{<}{\scriptstyle\sim}\>$}}

\newcommand{\msbar}{$\overline{\mbox{MS}}$ }

\begin{document}
\begin{titlepage}
\begin{flushright}
CERN-PH-TH/2004-175\\
September 2004
\end{flushright}
\par \vspace{10mm}
\begin{center}
{\Large \bf
Fragmentation in \boldmath{$H\to b\bar b$} Processes}
\end{center}
\par \vspace{2mm}
\begin{center}
{\large \bf G. Corcella}\\
\vspace{2mm}
{Department of Physics, CERN\\ 
Theory Division\\
CH-1211 Geneva 23, Switzerland}
\end{center}

\par \vspace{2mm}
\begin{center}
{\large \bf Abstract}
\end{center}
\begin{quote}
  \pretolerance 10000
I study bottom quark fragmentation 
in the Standard Model Higgs decay $H\to b\bar b$,
within the framework of perturbative fragmentation functions.
I resum large collinear logarithms $\ln(m_H^2/m_b^2)$ in the next-to-leading
logarithmic (NLL) approximation, using the DGLAP evolution
equations. Soft contributions to the \msbar coefficient function and to the
initial condition of the perturbative fragmentation function are 
resummed to NLL accuracy as well.
The implementation of collinear and soft resummation has a relevant 
impact on the $b$ energy spectrum, which exhibits a milder dependence
on factorization and renormalization scales and on the Higgs mass.
I present some predictions on the energy distribution of $b$-flavoured
hadrons in Higgs decay, making use of data from LEP and SLD experiments
to fit a few hadronization models.
I also compare the phenomenological results yielded by a few processes
recently provided with NLL collinear and soft resummations.
\end{quote}
\end{titlepage}
\section{Introduction}
The Higgs boson plays a crucial role in the Standard Model (SM)
of electroweak
interactions as it is responsible for the mechanism of mass generation.
However, this particle has not yet been experimentally discovered.

Searches for the Standard Model
Higgs boson have been performed at the LEP collider, 
are currently under way at the Tevatron, and will be ultimately
one of the main goals of experiments at the LHC. (see, for a review,
e.g. \cite{lh}).
In detail, the LEP experiments have set a lower bound on the Higgs mass at
$m_H>114.4$~GeV \cite{lep}, mainly using the
production channel $e^+e^-\to HZ$. 
The Tevatron will be able to 
exclude a Higgs boson with mass lower than 130 GeV within three
standard deviations \cite{tevatron}.
Future experiments at the LHC will be capable of going 
beyond and exploring the Higgs mass spectrum from 100 GeV to about 
1 TeV \cite{lhc}.

In order to accurately perform such searches, the use of 
precise QCD calculations
will be fundamental. In this paper, I consider the decay of the 
Standard Model Higgs
boson into $b\bar b$ pairs, i.e. $H\to b\bar b$.
In fact, the favourite discovery channel of
the Higgs at the Tevatron consists of processes where $H$ is produced in
association with a vector boson, i.e. $p\bar p\to VH$, where $V$ is
a $Z$ or a $W$, followed by the decays $H\to b\bar b$ and 
$V\to \ell_1\ell_2$, $\ell_1$ and $\ell_2$ being leptons.
At the LHC, the process $gg(q\bar q)\to H\to b\bar b$ will be affected
by large QCD backgrounds, which make the detection of
this decay channel more cumbersome.  
However, the process $H\to b\bar b$ will still play a role, 
in particular for $m_H\lsim 135$~GeV and Higgs production in 
association with $t\bar t$  pairs, i.e. $pp\to t\bar t H$
\cite{denegri}, with a $W$ boson \cite{denegri1},
in vector boson fusion \cite{ww}.

Hereafter, I shall address the issue of multiple gluon radiation in 
$H\to b\bar b$ processes. While fixed-order calculations are reliable
enough to predict total cross sections or widths, differential distributions
present terms, corresponding to collinear- or soft-parton radiation,
that need to be summed to all orders to obtain a reliable 
result. 

In particular, large mass logarithms $\ln(m_H^2/m_b^2)$,
which appear in the $b$-quark energy spectrum, can be resummed using
the approach of perturbative fragmentation functions \cite{mele},
which expresses the energy spectrum of a heavy quark as the 
convolution of a coefficient function, describing the emission of a 
massless parton, and a perturbative fragmentation function $D(m_b,\mu_F)$,
associated with the fragmentation of a massless parton into a massive
quark. The method of perturbative fragmentation can be used as long
as the heavy-quark mass $m$ is much smaller than the hard scale
of the process $Q$, i.e. $m\ll Q$. 
Given the current limits on the Higgs mass \cite{lep},
the perturbative fragmentation approach can certainly be used in
$H\to b\bar b$, since $m_b\ll m_H$.

The dependence of $D(m_b,\mu_F)$ on the factorization scale $\mu_F$
is determined by solving the Dokshitzer--Gribov--Altarelli--Parisi
(DGLAP) evolution equations \cite{ap,dgl}, once an initial condition at a scale
$\mu_{0F}$ is given. The universality of the initial condition
of the perturbative fragmentation function, first computed in
\cite{mele}, has been proved in a more general way in \cite{cc}.

Moreover, both coefficient function and initial condition of the 
perturbative fragmentation function contain terms that become 
large once the $b$ energy fraction $x_b$ gets close to 1.
Such terms correspond
to soft-gluon emission, and need to be resummed.
These contributions are process-dependent in the coefficient function and
process-independent in the initial condition of the perturbative
fragmentation function. The resummation of soft contributions to the
coefficient function of $H\to b\bar b$ will be investigated below.

Finally, in order to describe the $b$-quark non-perturbative  
fragmentation into $b$-flavoured mesons or baryons $B$, 
some phenomenological hadronization models can be used.
Relying on the
universality of the hadronization mechanism, we can tune
such models to data on $B$ production in $e^+e^-$ annihilation data from
LEP or SLD and use them to predict the 
$B$-hadron spectrum in Higgs decay.
Alternatively, we can use experimental data on the moments of 
the $B$ spectrum in $e^+e^-$ processes, fit the moments of the 
non-perturbative fragmentation
function and predict hadron-level moments in Higgs decay.

The plan of this paper is the following. In Section 2, I describe 
the calculation of the next-to-leading order (NLO) \msbar coefficient
function. 
The approach of perturbative fragmentation
and the resummation of collinear logarithms $\ln(m_H^2/m_b^2)$
will be discussed in Section 3.
Section 4 describes the implementation of soft resummation in
the coefficient function.
In Section 5, I shall present results on the $b$-quark energy spectrum in top
decay and investigate the effect of soft and collinear resummation.
In Section 6, hadron-level results in $x_B$ and $N$ spaces are presented, 
while Section 7 summarizes the main results and gives some concluding remarks.

\section{NLO coefficient function}

I consider Higgs decay into $b\bar b$ pairs at next-to-leading order
(NLO) in the strong coupling constant $\alpha_S$
\begin{equation}
H(p_H)\to b(p_b)\bar b(p_{\bar b}) \left(g(p_g)\right),
\end{equation}
and define the variables:
\begin{equation}
x_b={{2p_b\cdot p_H}\over{m_H^2}},\ \ \ 
x_g={{2p_g\cdot p_H}\over{m_H^2}}. 
\label{xbxg}
\end{equation}
The quantities $x_b$ and $x_g$ are the normalized
energy fractions of $b$ and $g$
in the Higgs rest frame. As they are expressed in the form of Lorentz-invariant
quantities, they can be computed in any frame, provided that 
the four components of the momenta of $b$, $g$ and $H$ are known.

In the framework of perturbative fragmentation functions, since
$m_b\ll m_H$, one can write
the differential width for the production of a
massive $b$ quark in Higgs decay via
the convolution:
\begin{eqnarray}
{1\over {\Gamma_0}} {{d\Gamma_b}\over{dx_b}} (x_b,m_H,m_b) &=&
\sum_i\int_{x_b}^1
{{{dz}\over z}\left[{1\over{\Gamma_0}}
{{d\hat\Gamma_i}\over {dz}}(z,m_H,\mu,\mu_F)
\right]^{\overline{\mathrm{MS}}}
D_i^{\overline{\mathrm{MS}}}\left({x_b\over z},\mu_F,m_b \right)} \nonumber \\
&&+ {\cal O}\left((m_b/m_H)^p\right) \; .
\label{pff}
\end{eqnarray}
In Eq.~(\ref{pff}), $d\hat\Gamma_i /dz$ is the differential width for the 
production of a massless parton $i$ in Higgs decay with 
an energy fraction $z$,
$D_i(x,\mu_F,m_b)$ is the perturbative
fragmentation function for a parton $i$ to fragment
into a massive $b$ quark, $\mu$ and
$\mu_F$ are the renormalization and factorization scales,
$\Gamma_0$ is the width of the Born process $H\to b\bar b$.
The term ${\cal O}\left((m_b/m_H)^p\right)$, with $p\geq 1$,
represents contributions which are power-suppressed for
$m_b\ll m_H$.

In this section I discuss the computation of the coefficient function
$(1/\Gamma_0) d\hat\Gamma_i/dz$, for the production of a massless $b$
in the \msbar factorization scheme. I shall neglect secondary $b\bar b$ 
production from gluon splitting and limit myself to considering the
perturbative fragmentation of a massless $b$ into a massive $b$.
In the summation on the right-hand side of Eq.~(\ref{pff}) I shall have
only the $i=b$ contribution.

I regularize ultraviolet, 
soft and collinear singularities in dimensional regularization, and define
the parameter $\epsilon$, which is related to the number of dimensions
$d$ via $d=4-2\epsilon$. 
In the computation of  
$d\hat\Gamma_b/dz$, care is to be taken about the treatment
of the Yukawa coupling of the $Hb\bar b$ vertex.
In fact, if the bare Yukawa coupling $y_b$ were used, the result would be
the following:
\begin{eqnarray}
{{d\hat\Gamma_b^{(0)}}\over{dz}} (z,m_H,\mu,\mu_r)&=&\Gamma_0
\left\{\delta(1-z)
+{{\alpha_S(\mu)}\over{2\pi}}
\Bigg[[P_{qq}(z)-3C_F\delta(1-z)]\right.\nonumber\\
&\times&
\left.\left.
\left(-{1\over\epsilon}+\gamma_E+\ln{{m_H^2}\over{4\pi\mu_r^2}}\right)
+G(z)\right]\right\}.
\label{masszero}
\end{eqnarray}
In Eq.~(\ref{masszero}), $\mu_r$ is the regularization scale,
remnant of the regularization procedure,
$G(z)$ is a function, independent of $\epsilon$ and $\mu_r$,
whose expression will be detailed later,
$\gamma_E=0.577\dots$ is the Euler constant, $C_F=4/3$, 
$P_{qq}(z)$ is the Altarelli--Parisi splitting function:
\begin{equation}
P_{qq}(z)=C_F\left({{1+z^2}\over{1-z}}\right)_+.
\end{equation}
Also, in Eq.~(\ref{masszero}) I have factorized the four-dimension
Born width:\footnote{In order
to get the correct finite term in Eq.~(\ref{masszero}),
one should compute the LO width to ${\cal O}(\epsilon)$ in 
dimensional regularization. 
The $d$-dimension width $\Gamma_d$ is related to $\Gamma_0$ by: 
$\Gamma_d=\Gamma_0(4\pi/m_H^2)^\epsilon[1+\epsilon
(2-\gamma)]$. Equations~(\ref{masszero}) and following
account for the correct finite term.}
\begin{equation}
\Gamma_0={3\over{16\pi}}m_Hy_b^2.
\label{born}
\end{equation}
Equation~(\ref{masszero}) shows that a pole $1/\epsilon$ is still present
in the differential width.
The contribution 
proportional to $P_{qq}(z)$ is associated with colliner radiation
from the massless $b$ quark and needs to be subtracted to give the
coefficient function. 
Analogous contributions have been found, e.g.
in the computation of the differential rate
of other processes such as $e^+e^-$
annihilation \cite{mele} or top decay \cite{cm}.
The additional term, where the pole $1/\epsilon$ multiplies
the quantity 
$\sim 3\alpha_S\ C_F\delta(1-z)$, 
instead has ultraviolet origin, and is characteristic of Higgs decay and
of the scalar nature of the coupling of the Higgs to quarks. 
In fact, unlike the vectorial current, which is conserved,  
the scalar current is anomalous.
Because of this extra term, if 
one naively calculated the total width integrating 
Eq.~(\ref{masszero}), one would still find
the $1/\epsilon$ pole, which is clearly unphysical.
\footnote{$P_{qq}(z)$ being an overall plus distribution, 
the integral of the term proportional to
$P_{qq}(z)$ over $z$ is clearly zero.}

In order to get a physical result, one must renormalize
the Yukawa coupling. 
In the \msbar renormalization scheme, the renormalized coupling
$\bar y_b(\mu)$ is related 
to $y_b$ via (see, for instance, 
the discussion in Appendix D of Ref.~\cite{maltoni}):
\begin{equation}
\bar y_b(\mu)=y_b\left\{1-{{\alpha_S(\mu)C_F}\over{4\pi}}
\left[4+3\left(-{1\over\epsilon}+\gamma_E+
\log{{m_H^2}\over{4\pi\mu^2}}\right)\right]+{\cal O}(\alpha_S^2)\right\}.
\label{ybyb}
\end{equation}
One can therefore reabsorb the
term $\sim 3\ \alpha_SC_F\delta(1-z)$ of Eq.~(\ref{masszero})
in the \msbar-renormalized coupling $\bar y_b(m_H)$
and evaluate the Born width in (\ref{born}) in terms
of $\bar y_b(m_H)$.

In the SM the Yukawa coupling is proportional to
the quark mass and, as discussed in \cite{maltoni},
Eq.~(\ref{ybyb}) is consistent with expressing 
$\bar y_b(m_H)$ in terms of the \msbar $b$ mass,
$\bar m_b(m_H)$:
\begin{equation}
\bar y_b(m_H)={{g\bar m_b(m_H)}\over{\sqrt{2}m_W}},
\label{yukms}
\end{equation}
where $m_W$ is the $W$ mass and $g$ is the coupling constant of $SU(2)$.

Furthermore, in Refs.~\cite{braaten,drees}, where the authors performed
the computation with a massive $b$ quark,  
it was shown that  
a large logarithm $\sim\alpha_S\ln(m_H^2/m_b^2)$ appears 
in the NLO total rate if one uses the $b$-quark pole mass in
the coupling.
Such a mass logarithm can be reabsorbed in the \msbar mass $\bar m_b(m_H)$,
which is related to the pole mass $m_b$ by:
\begin{equation}
\bar m_b(m_H)=m_b\left[1-{{\alpha_S(m_H)C_F}\over{4\pi}}\left(4+3
\ln{{m_H^2}\over{m_b^2}}\right)+{\cal O}(\alpha_S^2)\right].
\label{msmass}
\end{equation}
Subtracting the term $\sim P_{qq}(z)(-1/\epsilon+\gamma-\log 4\pi)$ 
from Eq.~(\ref{masszero}) and
giving an explicit expression to the function $G(z)$, 
one will get the \msbar $H\to b\bar b$ coefficient function:
\begin{eqnarray}
\left[{1\over{\bar\Gamma_0}} {{d\hat\Gamma_b}\over{dz}} (z,m_H,\mu,\mu_F)
\right]^{\overline{\mathrm{MS}}}
&=&
\delta(1-z)+{{\alpha_S (\mu)C_F}\over{2\pi}}
\left[\left({{1+z^2}\over{1-z}}\right)_+\ln{{m_H^2}\over{\mu_F^2}} 
\right.\nonumber\\
&+&
\left( {2\over 3} \pi^2
+ {3\over 2}\right) \delta(1-z)
+1-z-{3\over 2} {{z^2}\over{(1-z)_+}}\nonumber\\
&-&(1+z)[\ln(1-z)+2\ln z]
+6{{\ln z}\over{(1-z)_+}}\nonumber\\
&-&2{{\ln z}\over {1-z}}
+2\left.\left({{\ln(1-z)}\over{1-z}}\right)_+\right].
\label{coeffx}
\end{eqnarray}
In Eq.~(\ref{coeffx}) I have accounted for the renormalization
of the Yukawa coupling and denoted by $\bar\Gamma_0$ the
LO width in terms of $\bar y_b(m_H)$.
Also, in (\ref{coeffx}) the factorization scale $\mu_F$ 
will have to be taken of
the order of the Higgs mass, in such a way that the logarithm
$\ln(m_H^2/\mu_F^2)$ does not become too large.

In the following, I shall often make use of the \msbar coefficient function 
in Mellin moment space $\hat\Gamma_N$, which is defined by:
\begin{equation}
\hat\Gamma_N  =
\int_0^1 {dz  \ z^{N-1}
{1\over{\bar\Gamma_0}}{{d\hat\Gamma_b}\over{dz}}(z) }.
\end{equation}
In moment space Eq.~(\ref{coeffx}) reads:
\begin{eqnarray}
\hat\Gamma_N &=& 1+{{\alpha_S(\mu)C_F}\over{2\pi}}
\left\{\left[{1\over{N(N+1)}}
-2S_1(N)+{3\over 2}\right]
\ln{{m_H^2}\over{\mu_F^2}}\right.\nonumber\\
&+& {2\over 3}\pi^2+{3\over 2}+
{1\over N}-{1\over{N+1}}+{2\over{N^2}}+{2\over{(N+1)^2}}\nonumber\\
&-&4\psi_1(N)
+{1\over N}[\gamma+\psi_0(N+1)]+{1\over{N+1}}[\gamma+\psi_0(N+2)]
\nonumber\\
&+&{3\over 2} S_1(N+1)+S_1^2(N-1)+S_2(N-1)\Bigg\}
\label{coeffn}
\end{eqnarray}
In Eq.~(\ref{coeffn}), I have introduced the polygamma functions,
$\psi_k(x)$,
which are related to the Euler gamma function
$\Gamma(x)$ through:\footnote{Reference~\cite{cm} presents a typing mistake, 
since $\psi_k(x)$ is there defined as the $k$-th derivative of
$\Gamma(x)$. The numerical results of \cite{cm} 
are nonetheless correct.}
\begin{equation}
\psi_k(x)={{d^{k+1}\log\Gamma(x)}\over {dx^{k+1}}}.
\end{equation} 
Equation~(\ref{coeffn}) contains also the following combinations:
\begin{eqnarray}
S_1(N)&=&\psi_0(N+1)-\psi_0(1),\\
S_2(N)&=&-\psi_1(N+1)+\psi_1(1).
\end{eqnarray}
In moment space the convolution (\ref{pff}) can then be rewritten as
\begin{equation}
\Gamma_N(m_H,m_b) = \hat\Gamma_N(m_H,\mu,\mu_F) D_{b,N}(\mu_F,m_b),
\label{gamman}
\end{equation}
where  $\Gamma_N(m_H,m_b)$ and 
$D_{b,N}(\mu_F,m_b)$ are the moments of the massive 
differential rate and of the perturbative fragmentation function, respectively.

\section{Perturbative fragmentation\\ and collinear resummation}
The perturbative fragmentation function $D_b(x,\mu_F,m_b)$
introduced in Eq.~(\ref{pff}) expresses the transition of a massless $b$
into a massive $b$. 
Its value at any scale $\mu_F$ can be obtained by solving the DGLAP 
evolution equations
\cite{ap,dgl}, once an initial condition is given.
As shown in \cite{cc}, as long as contributions
proportional to powers of $(m_b/m_H)^p$ can be neglected,
the initial condition of the perturbative
fragmentation function at a scale $\mu_{0F}$
is process-independent. 
The NLO initial condition in the \msbar factorization scheme reads \cite{mele}:
\begin{equation}
D_b^{\rm ini}(x_b,\mu_{0F},m_b)=\delta(1-x_b)+
{{\alpha_S(\mu_0^2)C_F}\over{2\pi}}
\left[{{1+x_b^2}\over{1-x_b}}\left(\ln {{\mu_{0F}^2}\over{m_b^2}}-
2\ln (1-x_b)-1\right)\right]_+.
\label{dbb}
\end{equation}
The authors of Ref.~\cite{alex} have 
recently calculated $D_b^{\rm ini}(x,\mu_{0F},m_b)$
to next-to-next-to-leading order (NNLO), i.e.
up to ${\cal O}(\alpha_S^2)$.
For the purpose of this paper, where the coefficient function has been
calculated to NLO, the perturbative fragmentation
will be used to NLO as well.

The solution of the DGLAP equations in the non-singlet sector,
for the evolution from the scale
$\mu_{0F}$ to $\mu_F$, is given by:
\begin{eqnarray}
D_{b,N}(\mu_F,m_b)&=&
D_{b,N}^{\rm ini}(\mu_{0F},m_b)\exp\left\{ {{P_N^{(0)}}\over{2\pi b_0}}
\ln{{\alpha_S(\mu^2_{0F})}\over {\alpha_S(\mu^2_F)}}\right.\nonumber\\
&+&
\left.{{\alpha_S(\mu^2_{0F})-\alpha_S(\mu^2_F)}\over{4\pi^2b_0}}
\left[P_N^{(1)}-{{2\pi b_1}\over {b_0}}P_N^{(0)}\right]\right\}.
\label{dresum}
\end{eqnarray}
In Eq.~(\ref{dresum}), $D_{b,N}^{\rm ini}(\mu_{0F},m_b)$
is the $N$-space counterpart of Eq.~(\ref{dbb});
$P_N^{(0)}$ and $P_N^{(1)}$ are the Mellin transforms of the LO and
NLO Altarelli--Parisi splitting functions, and their
expression can be found in \cite{mele};
$b_0$ and $b_1$ are the first two coefficients of the QCD $\beta$-function
\begin{equation}
b_0={{33-2n_f}\over {12\pi}},\ \ b_1={{153-19n_f}\over{24\pi^2}},
\label{b0b1}
\end{equation}
which enter in the following expression for the strong coupling constant at
a scale $Q^2$:
\begin{equation}
\alpha_S(Q^2)={1\over {b_0\ln(Q^2/\Lambda^2)}}
\left\{ 1-{{b_1\ln\left[\ln (Q^2/\Lambda^2)\right]}\over
{b_0^2\ln(Q^2/\Lambda^2)}}\right\}.
\label{alpha}
\end{equation}
In (\ref{b0b1}), $n_f$ is the number of active flavours.
Equation~(\ref{dresum}) resums to all orders terms containing 
$\ln(\mu_F^2/\mu_{0F}^2)$.
In particular, leading (LL)
($\alpha_S^n\ln^n(\mu_F^2/\mu_{0F}^2)$)
and next-to-leading (NLL) ($\alpha_S^n
\ln^{n-1}(\mu_F^2/\mu_{0F}^2)$)  logarithms are resummed.
For an evolution from $\mu_{0F}\simeq m_b$ to $\mu_F\simeq m_H$, 
mass logarithms $\ln(m_H^2/m_b^2)$ 
are hence resummed to NLL accuracy (collinear resummation).
Moreover, setting $\mu_{0F}\simeq m_b$ in Eq.~(\ref{dbb}) prevents
the logarithm $\ln(\mu_{0F}^2/m_b^2)$ from getting too large.

If the calculation were performed with a massive 
$b$ quark, along the lines of Refs.~\cite{braaten,drees}, 
contributions $\sim\alpha_S P_{qq}(x_b)\ln(m_H^2/m_b^2)$, equivalent
to the collinear pole in massless approximation,
would be found in the $x_b$ differential spectrum.
Hence, using the DGLAP evolution equations allows 
the large mass logarithms appearing in the massive computation to be resummed.

Before closing this section, I wish to 
point out that, for the purpose of factorization and 
collinear resummation, using in $D_b^{\rm ini}(x_b,\mu_{0F},m_b)$
and in the DGLAP evolution equations 
the pole or the \msbar $b$ mass is not as essential as it is in the
Yukawa coupling (\ref{ybyb}).
In fact, both mass definitions lead to the same results
within the given LL or NLL logarithmic
accuracy. In the following, I shall assume that in Eq.~(\ref{dbb}) $m_b$
is the pole mass 
and will let $\mu_{0F}$ run in the range
$m_b/2<\mu_{0F}<2m_b$ in order to investigate the scale dependence of the 
prediction.
\footnote{If one wanted to use the
\msbar $b$ mass $\bar m_b(\mu_m)$, $\mu_m$ should be taken of the order of
$m_b$ rather than $m_H$. In fact, according to the factorization formula
(\ref{pff}), there is no dependence  on $m_H$ and on the hard-process
variables in the perturbative fragmentation function
$D_b(x_b,\mu_F,m_b)$.} The bottom pole mass has also been used
in the phenomenological analyses of Refs.~\cite{cc,cm,ccm},
within the framework of perturbative fragmentation functions.
 
\section{Soft resummation}
The \msbar coefficient function (\ref{coeffx}) and the initial
condition of the perturbative fragmentation function (\ref{dbb})
present terms 
$\sim 1/(1-x_b)_+$ and $\sim [\ln(1-x_b)/(1-x_b)]_+$ that become large
once $x_b\to 1$, which corresponds to soft-gluon emission.
In moment space, they correspond to single $\sim\ln N$ and double logarithms
$\sim\ln^2 N$ for large values of the Mellin variable $N$.
Such contributions are process-independent in the initial condition
of the perturbative fragmentation function, and have been resummed
in \cite{cc} in the NLL approximation.

In this section I would like 
to present the results for soft resummation in the
\msbar coefficient function.
First, it is instructive to write Eq.~(\ref{coeffn}) for large $N$:
\begin{eqnarray}
\hat\Gamma_N (m_H,\mu,\mu_F)&=& 1+{{\alpha_S(\mu)C_F}\over{2\pi}}
\left[\ln^2 N +\left( {3\over 2}+2\gamma-2\ln {{m_H^2}\over{\mu_F^2}}\right)
\ln N\right.\nonumber\\
&+& \left.K(m_H,\mu_F)+{\cal O}\left({1\over N}\right)\right],
\label{largen}
\end{eqnarray}
where $K(m_H,\mu_F)$ contains terms that are constant with
respect to $N$:
\begin{equation}
K(m_H,\mu_F)=\left({3\over 2}-2\gamma_E\right)\ln{{m_H^2}\over{\mu_F^2}}
+{5\over 6} \pi^2+{3\over 2} +{3\over 2} \gamma_E +\gamma_E^2.
\label{constants}
\end{equation}
Furthermore, to get Eq.~(\ref{largen}), I have used the large-$N$ 
expansions of the polygamma functions:
\begin{eqnarray}
\psi_0(N)&\sim& \ln N+{\cal O}\left({1\over N}\right),\\
\psi_1(N)&\sim& {\cal O}\left({1\over N}\right),\\
S_1(N)&\sim&\ln N+\gamma_E+{\cal O}\left({1\over N}\right),\\
S_2(N)&\sim& {{\pi^2}\over 6}+{\cal O}\left({1\over N}\right).
\end{eqnarray}
LL and NLL soft contributions to the \msbar coefficient function
can be resummed following standard methods \cite{ct,sterman}.

In particular, all the
steps that led to NLL resummation in the coefficient function for
$e^+e^-\to q\bar q$ processes in Ref.\cite{cc} can be repeated, thus
obtaining the resummed coefficient function for $H\to b\bar b$.
In fact, in both processes one has in the final state two
massless partons, as the $b$ and $\bar b$ are in the coefficient function,  
which are able to radiate soft- as well as
collinear-enhanced radiation.
The coefficients of $\ln N^2$ and $\ln N$ 
in Eq.~(\ref{largen}) are indeed the same as in the large-$N$ expansion of the 
$e^+e^-$ coefficient function \cite{cc}.

The resummed coefficient function can be obtained from the $e^+e^-$ one,
replacing the centre-of-mass energy squared $Q^2$ with $m_H^2$. One obtains:
\begin{equation}
\ln \Delta_N= \int_0^1 {dz {{z^{N-1}-1}\over{1-z}}}
\left\{\int_{\mu_F^2}^{m_H^2 (1-z)} {{dk^2}\over {k^2}}
A\left[\alpha_S(k^2)\right] +
{1\over 2} 
B\left[\alpha_S\left(m_H^2(1-z)\right)\right]\right\}.
\label{delta}
\end{equation}
In Eq.~(\ref{delta}), the two integration
variables are $z=1-x_g$ and $k^2=(p_b+p_g)^2(1-z)$, 
as in \cite{ct}. In soft approximation, $z\simeq x_b$;
for small-angle radiation, $k^2\simeq k_T^2$, the gluon transverse momentum
with respect to the $b$-quark direction.

The function $B(\alpha_S)$ is associated with the radiation emitted by 
the unobserved massless parton, namely the $\bar b$ if one observes the $b$. 
The argument of $B(\alpha_S)$ is the invariant mass of the unobserved jet,
i.e. $(p_{\bar b}+p_g)^2\simeq m_H^2(1-z)$ in soft approximation.

Moreover, Eq.~(\ref{delta}) is formally equal to 
the resummed \msbar coefficient function in Drell--Yan and deep inelastic
scattering (DIS)
with light quarks \cite{cmw}. For processes with massive quarks,
such as top quark decay \cite{ccm} or heavy quark production in
DIS \cite{cm1}, recently provided with NLL soft resummation,
the function $B(\alpha_S)/2$ should be replaced by a different one,
called $S(\alpha_S)$ in Refs.~\cite{ccm} and \cite{cm1}, which is 
characteristic of processes with heavy quarks and expresses soft radiation,
which is not collinear-enhanced.
The function $A(\alpha_S)$ in Eq.~(\ref{delta}) 
can be expanded as a series in $\alpha_S$ as:
\begin{equation}
A(\alpha_S)=\sum_{n=1}^{\infty}\left({{\alpha_S}\over
{\pi}}\right)^n A^{(n)}.
\end{equation}
The first two coefficients are mandatory to resum the 
coefficient function
to NLL accuracy \cite{ct}: 
\begin{equation}
A^{(1)}=C_F,
\end{equation}
\begin{equation}
A^{(2)}= {1\over 2} C_F \left[ C_A\left(
{{67}\over{18}}-{{\pi^2}\over 6}\right) -{5\over 9}n_f\right],
\end{equation}
where $C_A = 3$.
Likewise, the function $B(\alpha_S)$ can be expanded: 
\begin{equation}
B(\alpha_S)=\sum_{n=1}^{\infty}\left({{\alpha_S}\over
{\pi}}\right)^n B^{(n)}
\end{equation}
and, to NLL level, only the first term of the expansion is kept:
\begin{equation}
B^{(1)}=-{3\over 2}C_F.
\end{equation} 
The integral over $z$ can be performed making use of the 
replacement \cite{ct}:
\begin{equation}
z^{N-1}-1\to -\Theta\left( 1-{{e^{-\gamma_E}}\over N}-z\right) ,
\end{equation}
where $\Theta$ is the Heaviside step function. The function
$\Delta_N$ can be expressed in the usual form:
\begin{equation}
\Delta_N(m_H,\mu,\mu_F)=\exp\left[\ln N
g^{(1)}(\lambda)+ g^{(2)}(\lambda,\mu,\mu_F)\right]\, ,
\label{deltaint}
\end{equation}
with
\begin{equation}
\lambda=b_0\alpha_S(\mu)\ln N.
\end{equation}
In Eq.~(\ref{deltaint}) the term $\ln N g^{(1)}(\lambda)$ accounts
for the resummation of the leading logarithms $\alpha_S^n\ln^{n+1}N$
in the Sudakov exponent, and the function $g^{(2)}(\lambda,\mu,\mu_F)$
resums NLL terms $\alpha_S^n\ln^nN$.
Functions $g^{(1)}$ and $g^{(2)}$ can be obtained by simply setting
$Q^2= m_H^2$ in Eqs.~(34) and (35) of Ref.~\cite{cc}; 
the result is not here reported for the sake of brevity.

Following \cite{cc,cm1,cm}, I include in the Sudakov-resummed
coefficient function the constant terms $K(m_H,\mu_F)$ defined in
Eq.~(\ref{constants}) and obtain:
\begin{eqnarray}
\Delta_{N}^S(m_H,\mu,\mu_F)&=&
\left[ 1+{{\alpha_S(\mu)C_F}\over{2\pi}} K(m_H,\mu_F)\right]
\nonumber\\
&\times& \exp\left[\ln N
g^{(1)}(\lambda)+ g^{(2)}(\lambda,\mu,\mu_F)\right].
\label{sud}
\end{eqnarray}
Finally, the resummed result is matched to the fixed-order one, which will
yield a prediction valid at $x_b<1$ as well. The NLO result is added
to Eq.~(\ref{sud}) and, in order to avoid double counting, 
the ${\cal O}(\alpha_S)$ term of the resummed result is subtracted:
\begin{eqnarray}
\hat\Gamma_N^{\mathrm{res}}(m_H,\mu,\mu_F)
&=&\hat\Gamma_N^S(m_H,\mu,\mu_F)-
\left[\hat\Gamma_N^S(m_H,\mu,\mu_F)
\right]_{\alpha_S}\nonumber\\
&+&\left[\hat\Gamma_N(m_H,\mu,\mu_F)\right]_{\alpha_S},
\end{eqnarray}
where $[\hat\Gamma_N^S]_{\alpha_S}$ and
$[\hat\Gamma_N]_{\alpha_S}$ are respectively the expansion of
Eq.~(\ref{delta}) up to ${\cal O}(\alpha_S)$ and the full fixed-order
coefficient function at ${\cal O} (\alpha_S)$ (\ref{coeffn}).

\section{Bottom-quark energy distribution}
In this section I present the energy spectrum of bottom quarks in Higgs decay,
according to the calculation above described.
I shall investigate the phenomenological effect of collinear and
soft resummation, and the dependence of the prediction on factorization
and renormalization scales.
The results, which have been presented in analytic form in Mellin space,
will be inverted numerically to $x_b$ space choosing the
integration contour according to the
minimal prescription \cite{cmnt}.
I shall plot the normalized rate $(1/\Gamma)(d\Gamma/dx_b)$, where
$\Gamma$ is the NLO $H\to b\bar b$ width, calculated in 
Refs.~\cite{braaten,drees}. 
I shall set $m_H=120$~GeV, $m_b=5$~GeV, $n_f=5$, 
$\Lambda=200$~MeV. 

\begin{figure}
\centerline{\resizebox{0.65\textwidth}{!}{\includegraphics{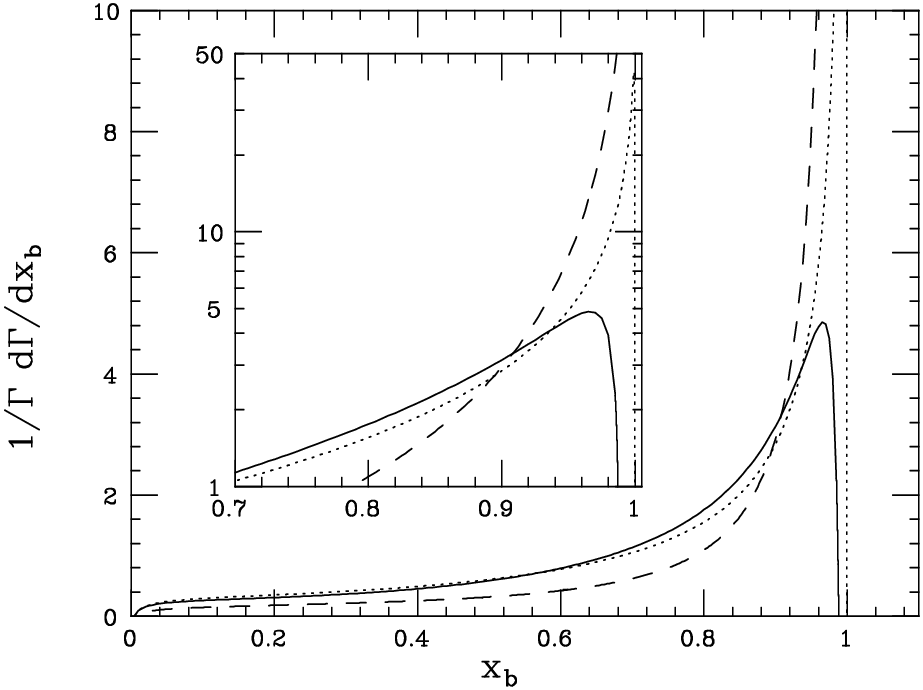}}}
\caption{$b$-quark energy distribution in Higgs decay according to
the NLO massive calculation (dashed line), including NLL collinear resummation
(dots) and both NLL collinear and soft resummations (solid).   
I have set: $\mu=\mu_F=m_H$, $\mu_0=\mu_{0F}=m_b$.
In the inset figure, 
the same curves are shown at large $x_b$ and on a logarithmic
scale.}
\label{enb}
\end{figure}

In Fig.~\ref{enb} I have plotted the energy spectrum of the $b$-quark,
according to the NLO massive 
calculation (dashed line), which can be obtained from Eq.~(\ref{pff}) without
evolving the perturbative fragmentation function,
including collinear resummation (dotted line) and with both collinear and
soft resummations (solid). I have set 
$\mu=\mu_F=m_H$, $\mu_0=\mu_{0F}=m_b$.

We can notice a remarkable effect of both collinear and soft resummations.
The fixed-order calculation lies below the resummed ones and 
grows as $x_b\to 1$ because of a behaviour $\sim 1/(1-x_b)$.
The collinear-resummed spectrum exhibits instead a sharp peak at large $x_b$;
after both soft and collinear resummations, the distribution is
further smoothed and presents the Sudakov peak at $x_b\simeq 0.97$.

Although one implements both collinear and soft resummations, 
the prediction is still not reliable
at very small and large $x_b$. In fact, even the resummed distributions 
become negative
once $x_b$ approaches 0 or 1. At small $x_b$, the coefficient function contains
a term $\sim\log x_b$ which has not been resummed yet; 
at very large $x_b$, one starts to get sensitive to missing non-perturbative
power corrections. The range of reliability of a purely perturbative
computation is typically $x_b\lsim 1-\Lambda/m_b$ \cite{cc}.

\begin{figure}
\centerline{\resizebox{0.65\textwidth}{!}{\includegraphics{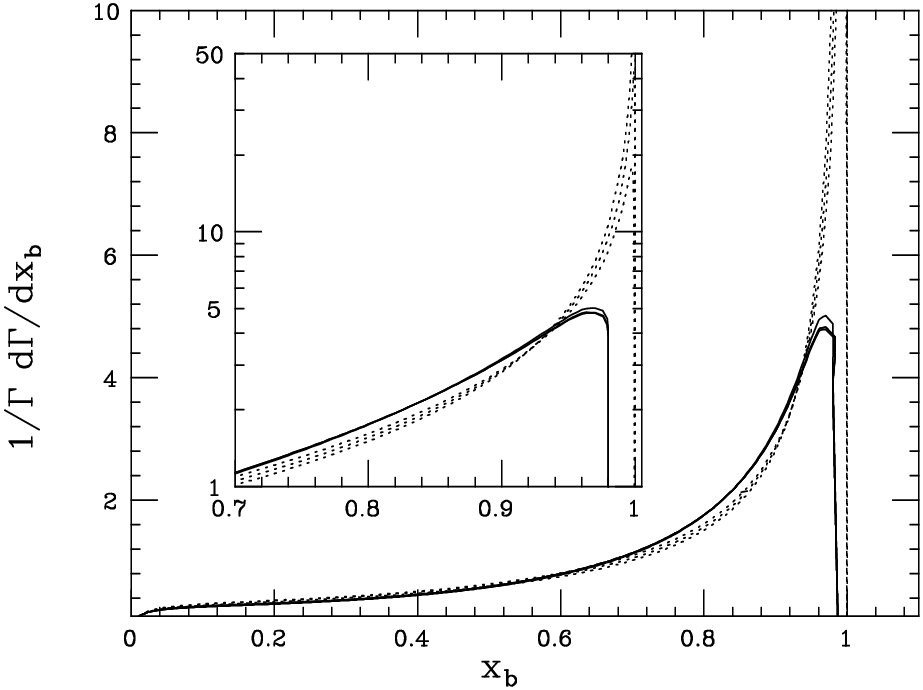}}}
\caption{Dependence of the $b$ spectrum on the factorization scale $\mu_F$,
with (solid lines) and without (dots) NLL soft resummation. 
The other scales are fixed at $\mu=m_H$ and $\mu_0=\mu_{0F}=m_b$.}
\label{hfac}
\end{figure}

\begin{figure}
\centerline{\resizebox{0.65\textwidth}{!}{\includegraphics{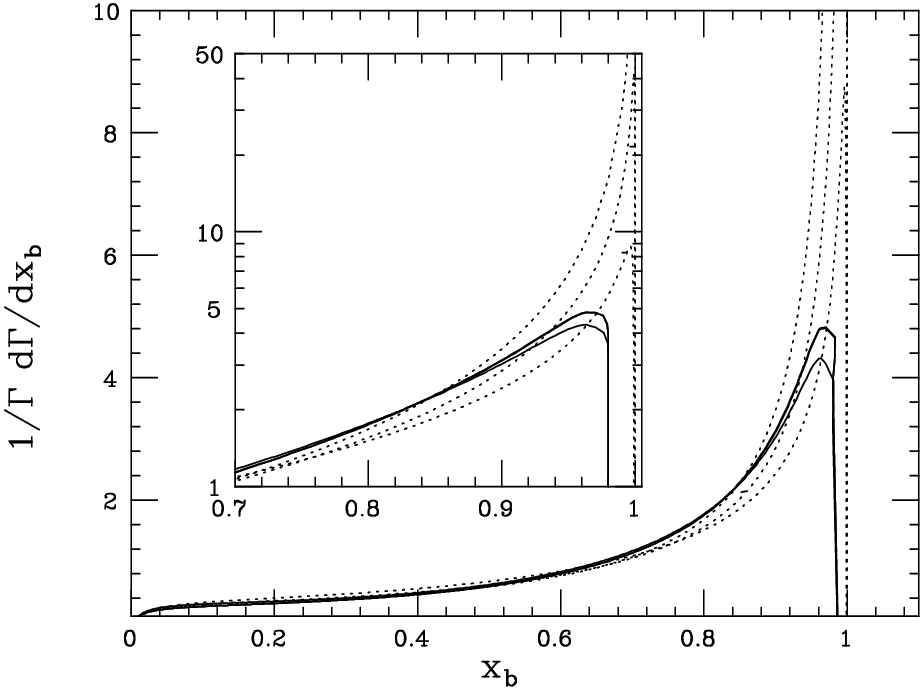}}}
\caption{As in Fig.~\ref{hfac}, but for different values of 
the factorization scale $\mu_{0F}$.
The other scales are fixed at $\mu=\mu_F=m_H$ and $\mu_0=m_b$.}
\label{hfac0}
\end{figure}

\begin{figure}
\centerline{\resizebox{0.65\textwidth}{!}{\includegraphics{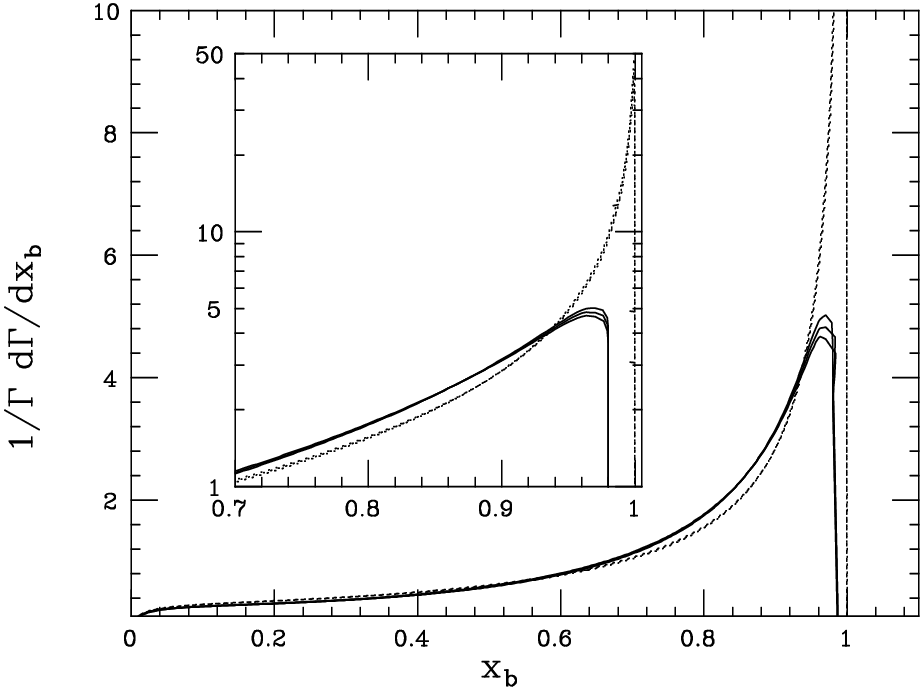}}}
\caption{As in Fig.~\ref{hfac}, but varying the renormalization scale
$\mu$. The other scales are fixed at $\mu_F=m_H$ and $\mu_{0F}=\mu_0=m_b$.}
\label{hren}
\end{figure}

\begin{figure}
\centerline{\resizebox{0.65\textwidth}{!}{\includegraphics{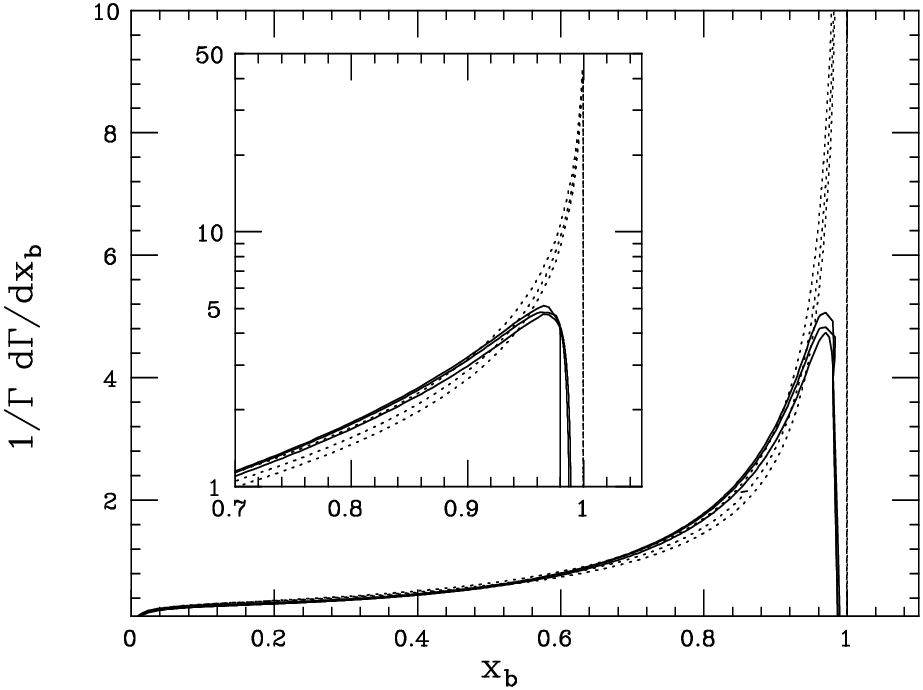}}}
\caption{As in Fig.~\ref{hren}, but for different
values of the renormalization scale $\mu_0$.
The other scales are fixed at $\mu=\mu_F=m_H$ and $\mu_{0F}=m_b$.}
\label{hren0}
\end{figure}

In Figs.~\ref{hfac}-\ref{hren0}, I show the dependence of the prediction
on the factorization and renormalization scales that appear in the
calculation, i.e. $\mu$ and $\mu_F$ in Eq.~(\ref{coeffx}),
$\mu_0$ and $\mu_{0F}$ in (\ref{dbb}).
I let $\mu$ and $\mu_F$ assume the values $m_H/2$, $m_H$ and
$2m_H$; as for the scales $\mu_0$ and $\mu_{0F}$, as discussed 
in Section 3, they will be taken equal to $m_b/2$, $m_b$ and $2m_b$.
The dependence on the scales is logarithmic, hence we expect it to be
more visible once the scales vary around small values, i.e.
around $m_b$ rather than $m_H$.

Collinear resummation is turned on in all figures, but plots are with 
or without soft-gluon resummation.
The general feature of these results is that after
both collinear- and soft-enhanced terms are resummed, the final prediction
exhibits very little scale dependence.
In particular, we note that the role of soft resummation is 
crucial to weaken the dependence on scales $\mu_F$ (Fig.~\ref{hfac}), 
$\mu_0$ (Fig.~\ref{hren0}) and especially $\mu_{0F}$ (Fig.~\ref{hfac0}).
The dependence on the renormalization scale $\mu$ is instead very weak
even resumming only collinear logarithms, as shown in Fig.~\ref{hren}.
A milder sensitivity to renormalization and 
factorization scales corresponds to a reduction of
the theoretical uncertainty on the result and is a meaningful effect
of collinear and soft resummations.

\begin{figure}
\centerline{\resizebox{0.65\textwidth}{!}{\includegraphics{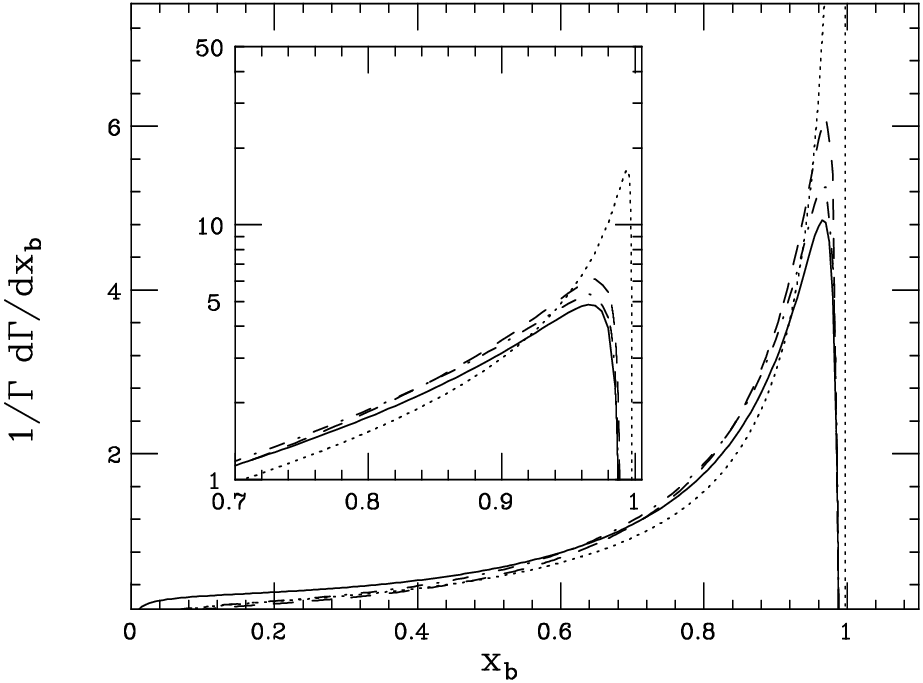}}}
\caption{$b$-quark energy distribution in Higgs decay (solid), 
top decay (dots) and $e^+e^-\to b\bar b$ processes at $\sqrt{s}=91.2$~GeV
(dashes) and 120 GeV (dot-dashes). 
All predictions are given by a NLO calculation provided with NLL
collinear and soft resummations. In Higgs decay I have set 
$\mu=\mu_F=m_H$, in top decay  
$\mu=\mu_F=m_t=$~175 GeV, in $e^+e^-$ annihilation 
$\mu=\mu_F=\sqrt{s}$.
All plots are for $\mu_0=\mu_{0F}=m_b$.}
\label{hzt}
\end{figure}

Moreover, I would like to compare the $b$-quark energy
distribution in three processes that have been investigated within
the framework of perturbative fragmentation functions and provided with
NLL collinear and soft resummations, namely
$e^+e^-\to b\bar b$ \cite{cc}, top decay $t\to bW$ \cite{cm,ccm}
and, in this paper, Higgs decay.
Because of the universality of the fragmentation process, possible differences
in such spectra will be related to the different coefficient functions
and mass scales involved.
Furthermore, I shall
consider $e^+e^-$ at $\sqrt{s}=91.2$~GeV, the centre-of-mass energy
of LEP I and SLD, and, in order to test the effect of the coupling,
vectorial in $e^+e^-\to b\bar b$ and scalar in $H\to b\bar b$,
also at $\sqrt{s}=$~120~GeV, the default Higgs-mass value 
throughout this paper. In top decay I shall set $m_t=$~175 GeV and
$m_W=$~80 GeV.
From Fig.~\ref{hzt} one learns that the shapes of the three distributions
exhibit some differences. 
The $b$ spectrum in $H$ decay is the highest at small $x_b$ and the
lowest at large $x_b$;
in top decay it is shifted toward large $x_b$ and peaked very close to 1.
The $e^+e^-\to b\bar b$ prediction lies within the other two at small
and very large $x_b$.
The effect of different values of $\sqrt{s}$ in the
$e^+e^-$ process is visible mainly around the Sudakov peak;
setting $\sqrt{s}=m_H$ makes the spectrum more similar to the 
Higgs decay one.

\begin{figure}
\centerline{\resizebox{0.65\textwidth}{!}{\includegraphics{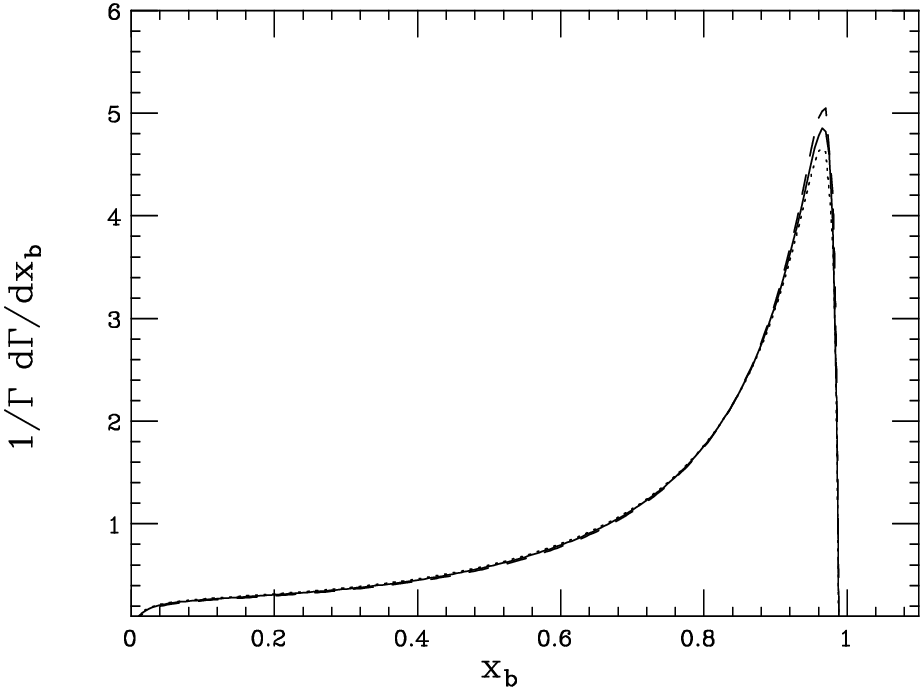}}}
\caption{$b$-quark energy distribution in Higgs decay, 
including NLL collinear and soft resummations, for a Higgs mass of
110 GeV (dashes), 120 GeV (solid) and 130 GeV (dots).}
\label{hmass}
\end{figure}

Before closing this section, 
it is interesting to investigate the dependence of the resummed
prediction on the Higgs mass. In Fig.~\ref{hmass}, we plot the
NLL collinear- and soft-resummed spectrum for $m_H=110$, 120 and 130 GeV.
We note that the spectra change very little with the
chosen value of $m_H$, and some small effect is only visible around
the Sudakov peak.

\section{Hadron-level results}
I shall now present results on the energy spectrum of $b$-flavoured
hadrons $B$ in Higgs decay. Similarly to the parton-level analysis, 
for a hadron of momentum $p_B$ 
one defines the $B$ normalized energy fraction in the Higgs rest frame:
\begin{equation}
x_B={{2p_B\cdot p_H}\over{m_H^2}}.
\end{equation}
Up to power corrections, one writes the hadron-level spectrum as 
the convolution of the parton-level one, calculated above,
with the non-perturbative 
fragmentation function $D^{np}(x)$, associated with the hadronization of the  
$b$ quark into the $B$ hadron:
\begin{equation}
{1\over {\Gamma}} {{d\Gamma_B}\over{dx_B}} (x_B,m_H,m_b)={1\over{\Gamma}}
\int_{x_B}^1 {{{dz}\over z}{{d\Gamma_b}\over {dz}}(z,m_H,m_b)
D^{np}\left({x_B\over z}\right)}.
\label{npff}
\end{equation}
The non-perturbative fragmentation function is to be extracted from 
experimental data. As the Higgs particle has not been discovered yet, 
one obviously does not have any data on $B$ production in Higgs decay.
However, relying on the universality of the hadronization transition,
one can fit some hadronization models to data on $B$ production in
$e^+e^-$ experiments and use them to predict the $B$ spectrum in
$H\to b\bar b$ processes.
In order for the fitting procedure to be consistent, the 
$e^+e^-\to b\bar b$ perturbative process 
is to be described as done for Higgs decay.
I shall have to use NLO coefficient functions, NLL collinear and soft 
resummation, and consistent values for the scales appearing in the
calculation, e.g. $\mu=\sqrt{s}$ if I set $\mu=m_H$ in
$H\to b\bar b$.
In fact, it was recently found \cite{canas,mlm}
that such a consistency is crucial if one wishes to make use
of non-perturbative information taken from $e^+e^-$ data to describe
the data on
$B$-hadron production at hadron colliders, such as the Tevatron.

In this paper, as for the non-perturbative
fragmentation function, 
I shall consider the following three hadronization models:
a power law with two parameters
\begin{equation}
D^{np}(x;\alpha,\beta)={1\over{B(\beta +1,\alpha +1)}}(1-x)^\alpha x^\beta,
\label{ab}
\end{equation}
the model of Kartvelishvili et al. \cite{kart}
\begin{equation}
D^{np}(x;\delta)=(1+\delta)(2+\delta) (1-x) x^\delta
\label{kk}
\end{equation}
and the Peterson model \cite{peterson}
\begin{equation}
D^{np}(x;\epsilon)={A\over {x[1-1/x-\epsilon/(1-x)]^2}}.
\label{peter}
\end{equation}
In Eq.~(\ref{ab}), $B(x,y)$ is the Euler beta function; in
(\ref{peter}) $A$ is a normalization constant.
The parameters $\alpha$, $\beta$, $\delta$ and $\epsilon$ are to be extracted 
from experimental data.
In Ref.\cite{ccm}, such models have been fitted to ALEPH \cite{aleph} data 
on $B$ mesons and it was found that models (\ref{ab}) and (\ref{kk}) 
yield very good fits, while the Peterson model is marginally
consistent.
In Ref.~\cite{cor}, also the SLD data \cite{sld} 
were considered and it was found that
Eqs.~(\ref{ab}) and (\ref{kk}) lead to good fits, 
and Eq.~(\ref{peter}) is instead
unable to reproduce the SLD data. Moreover, using a soft-resummed
perturbative calculation was essential to describe the SLD data.
However, discrepancies were found  in Ref.~\cite{cor}
between the best-fit parameters
$\alpha$, $\beta$ and $\delta$, according to whether one fits the 
models to ALEPH or SLD.
Unlike ALEPH, the SLD data contain some $b$-flavoured baryons, mainly the 
$\Lambda_b$, but it is a pretty small fraction of the whole
sample; hence, it may not be correct to conclude that the differences 
reported in \cite{cor} are due to the baryons reconstructed in
SLD. Moreover, detailed analyses and comparisons 
on $b$-fragmentation in the four
LEP experiments and SLD are currently missing. 

Here I shall try instead a combined fit of both ALEPH and SLD data samples
and investigate whether one is able to find a suitable parametrization 
of the hadronization models (\ref{ab})--(\ref{peter})
that would be consistent with both experiments. 
As in Refs.~\cite{cm,ccm,cor}, 
I shall consider data for $0.18\lsim x_B\lsim 0.94$,
so as to avoid data points close to 0 or 1, where the presented
calculation is not reliable. Furthermore, when doing the fits,
I shall neglect the correlations among data points and
sum statistical and systematic errors in quadrature.

\begin{table}[ht]
\begin{center}
\begin{tabular}{||c|c||}\hline
$\alpha$&$0.90\pm 0.15$  \\
\hline
$\beta$&$16.23\pm 1.37$  \\
 \hline
$\chi^2(\alpha,\beta)$/dof&33.42/31 \\
\hline
$\delta$&$17.07\pm 0.39 $ \\
\hline
$\chi^2(\delta)$/dof&33.80/32  \\
\hline
$\epsilon$&$(1.71\pm 0.09)\times 10^{-3}$\\
\hline
$\chi^2(\epsilon)$/dof&166.36/32\\
 \hline
\end{tabular}
\end{center}
\caption{\label{tabx}
Results of combined
fits to $e^+e^-\to b\bar b$ data from ALEPH and SLD collaborations, 
using NLO coefficient functions,
NLL DGLAP evolution and NLL soft-gluon
resummation.
I have set $\Lambda=200$~MeV, $\mu_{0F}=\mu_0=m_b=5$~GeV and
$\mu_F=\mu=\sqrt{s}=91.2$~GeV. $\alpha$ and $\beta$ are the parameters in
the power law (\ref{ab}), $\delta$ refers to (\ref{kk}), $\epsilon$
to (\ref{peter}).}
\end{table}
\begin{table}
\begin{center}
\begin{tabular}{||c|c||}\hline
$\alpha$&$1.01\pm 0.12$  \\
\hline
$\beta$&$14.73\pm 0.94$  \\
 \hline
$\chi^2(\alpha,\beta)$/dof&64.19/31 \\
\hline
$\delta$&$14.67\pm 0.30 $ \\
\hline
$\chi^2(\delta)$/dof&64.20/32  \\
\hline
$\epsilon$&$(2.76\pm 0.14)\times 10^{-3}$\\
\hline
$\chi^2(\epsilon)$/dof&287.55/32\\
 \hline
\end{tabular}
\end{center}
\caption{\label{tabxno}
As in Table~\ref{tabx}, but without soft resummation in the 
parton-level calculation.}
\end{table}
In Table~\ref{tabx} 
the best-fit parameters are quoted, along with the $\chi^2$ per
degree of freedom.
Both power law and Kartvelishvili models fit the data quite well, while
the Peterson non-perturbative model is unable to describe them.
It is also interesting to investigate whether the implementation
of soft resummation has an impact on the fit.
Table~\ref{tabxno} shows the best-fit parameters obtained
without resumming soft-gluon contributions to the perturbative calculation of
$e^+e^-\to b\bar b$.
It should be noted 
that the fit gets worse and no hadronization model is capable
of yielding a reasonably small $\chi^2/\mathrm{dof}$.

\begin{figure}
\centerline{\resizebox{0.65\textwidth}{!}{\includegraphics{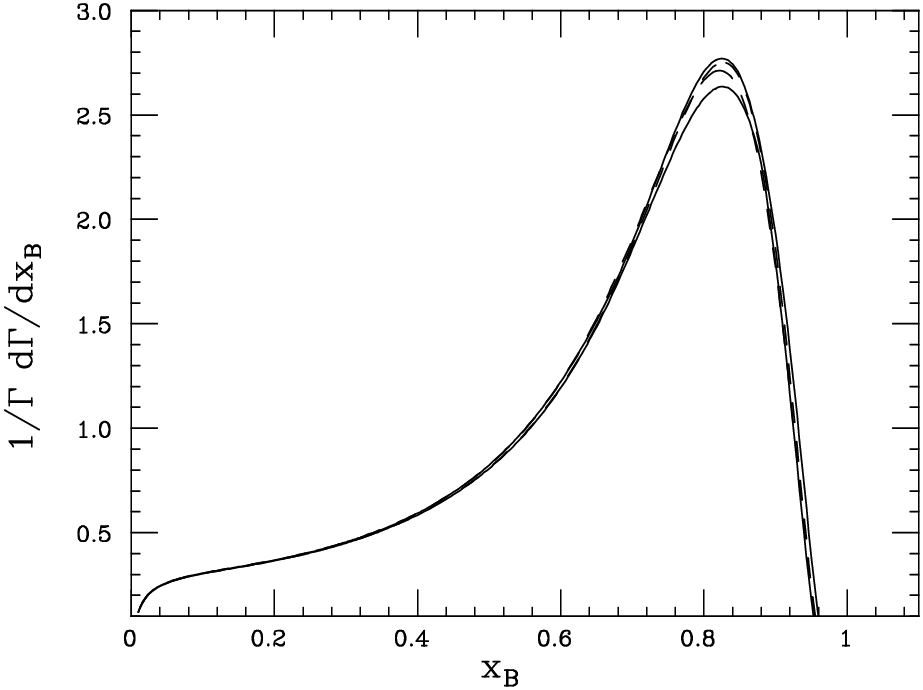}}}
\caption{$B$-hadron spectrum in Higgs decay, modelling the hadronization
according to the power law (\ref{ab}) (solid) and the Kartvelishvili
model (\ref{kk}) (dashes).
Plotted are the edges of bands at one-standard-deviation
confidence level for the non-perturbative parameters $\alpha$, $\beta$ and
$\delta$, as reported in Table~\ref{tabx}.
In the perturbative calculation, the scales have been set to:
$\mu=\mu_F=m_H=120$~GeV, $\mu_0=\mu_{0F}=m_b=5$~GeV.}
\label{highad}
\end{figure}

In Fig.~\ref{highad} I present a prediction on the $x_B$ distribution in
Higgs decay, using the NLL collinear- and soft-resummed 
perturbative calculation, and the best-fit parameters of
models (\ref{ab}) and (\ref{kk}) quoted in Table~\ref{tabx}.
I discard the Peterson model as it does not acceptably describe
the considered data sample. In order to account for the errors on the
best-fit parameters given in Table~\ref{tabx}, for each 
model I plot a band corresponding to
a prediction at one-standard-deviation confidence level
for the fitted non-perturbative parameters $\alpha$, $\beta$ and $\delta$.

\begin{figure}
\centerline{\resizebox{0.65\textwidth}{!}{\includegraphics{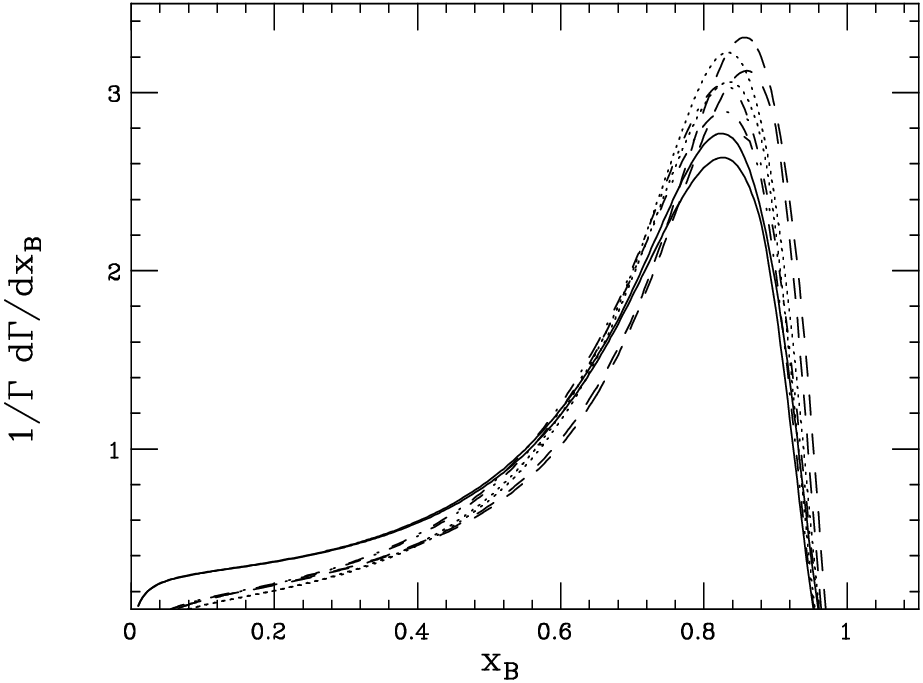}}}
\caption{$B$-hadron spectra in Higgs decay (solid), in top decay (dashes)
and in $e^+e^-$ annihilation at $\sqrt{s}=91.2$~GeV (dotted) and 
120 GeV (dot-dashed), according to the hadronization model
(\ref{ab}), at one-standard-deviation confidence level for the parameters 
$\alpha$ and $\beta$. Factorization and renormalization
scales are chosen as in Fig.~\ref{hzt}.}
\label{hzthad}
\end{figure}

Fig.~\ref{highad} shows that the two predictions yielded
by models (\ref{ab}) and (\ref{kk}) are statistically consistent.
In fact, from Table~\ref{tabx} one learns 
that, within the error range, the parameter $\alpha$ of Eq.~(\ref{ab})
is consistent with 1 and $\beta$ of Eq.~(\ref{ab}) is consistent with
$\delta$ of Eq.~(\ref{kk}). It is therefore reasonable that the predictions of
the power law with two parameters and of the Kartvelishvili model
agree.
\\
\indent As I did for the analysis at parton level, I present 
in Fig.~\ref{hzthad} the $B$-hadron
spectra in three different processes, i.e. $e^+e^-$
annihilation, Higgs and top decay, using everywhere 
a NLO and NLL resummed perturbative
calculation and the hadronization model (\ref{ab}), with the best-fit
parameters reported in Table~\ref{tabx}.
As in the parton-level 
analysis, I make consistent choices for the scales involved and
consider two centre-of-mass energies for
the electron--positron process, i.e. $\sqrt{s}=91.2$ and 120 GeV.
For each process I plot a band corresponding to a prediction at
one-standard-deviation confidence level for the parameters
$\alpha$ and $\beta$ of Eq.~(\ref{ab}).

Figure~\ref{hzthad} shows that the three spectra are statistically different 
and that the result is pretty similar to the one already found at parton-level
in Fig.~\ref{hzt}, which is reasonable as we are convoluting the
$x_b$ distribution with the same non-perturbative fragmentation
function.
Setting $\sqrt{s}=m_H$ makes the $e^+e^-\to b\bar b$ spectrum closer to the
$H$-decay one, especially at middle and large values of $x_B$, 
though meaningful differences are still present
at small $x_B$ and around the peak.
\\ \indent
Before closing this section, I would like to present results in moment space,
using the experimental moments on $B$ production in $e^+e^-$ annihilation
from the DELPHI Collaboration \cite{delphi}, along the lines of \cite{ccm}.
The advantage of working in moment space, as suggested in \cite{canas},
is that one does not need to rely on any specific functional form
of the non-perturbative fragmentation function.
\begin{table}
\small
\begin{tabular}{| c | c c c c |}
\hline
& $\langle x\rangle$ & $\langle x^2\rangle$ & $\langle x^3\rangle$
& $\langle x^4\rangle$ \\
\hline
$e^+e^-$ data $\sigma_N^B$&0.7153$\pm$0.0052 &0.5401$\pm$0.0064 &
0.4236$\pm$0.0065 &0.3406$\pm$0.0064  \\
\hline
\hline
$e^+e^-$ NLL $\sigma_N^b$   & 0.7801 & 0.6436 & 0.5479 & 0.4755  \\
\hline
$D^{np}_N$ & 0.9169 & 0.8392 & 0.7731 & 0.7163 \\
\hline
\hline
$H$-decay NLL $\Gamma^b_N$ & 0.7578 & 0.6162 & 0.5193 & 0.4473  \\
\hline
$H$-decay $\Gamma^B_N$ & 0.6948 & 0.5171 & 0.4015 & 0.3204 \\
\hline
\end{tabular}
\caption{\label{tablen}\small  Experimental data for the moments
$\sigma^B_N$ from
DELPHI~\protect\cite{delphi}, the resummed $e^+e^-$ perturbative
calculations for $\sigma^b_N$~\protect\cite{cc},
the extracted non-perturbative contribution
$D^{np}_N$. Using the resummed 
perturbative result $\Gamma_N^b$, a prediction for
the moments $\Gamma^B_N$ in $H\to b\bar b$ processes is given.
The experimental
error should be propagated to the final prediction.}
\end{table}
As done in \cite{ccm}, I have quoted in Table~\ref{tablen}  
the first four moments on $B$-production at DELPHI, the computed moments of
$e^+e^-$ and Higgs decay perturbative calculations, the extracted moments
of the non-perturbative fragmentation function and the predictions
for the moments of the $B$ spectra in $H\to b\bar b$.

In moment space convolutions are turned into ordinary products, 
and we therefore have: $\sigma_N^B=\sigma_n^b D^{np}_N$,
$\Gamma_N^B=\Gamma^b_N D^{np}_N =\Gamma^B_N\sigma_N^B/\sigma_N^b$,
where $\sigma_N$ and $\Gamma_N$ are the moments of 
$e^+e^-$-annihilation cross section and Higgs-decay width 
at hadron or parton level.
Comparing Table~\ref{tablen} with the results 
of \cite{ccm}, it should be noted that all considered
moments in $H$ decay are lower than those in
$e^+e^-$ processes and in top decay.
This result is consistent with the spectra in $x_B$-space presented
in Fig.~\ref{hzthad}.

\section{Conclusions}
I have considered bottom-quark fragmentation in Standard Model
Higgs decay 
$H\to b\bar b$ within the approach of perturbative fragmentation functions.
I have computed the \msbar NLO coefficient function and resummed collinear
logarithms $\ln (m_H^2/m_b^2)$ by using the DGLAP evolution equations 
in the NLL approximation. In the calculation of the coefficient
function, the use of the \msbar-renormalized
Yukawa coupling turned out to be essential.

I have resummed NLL soft terms in the coefficient function and
matched the resummed result with the exact NLO one.
Soft resummation in the coefficient function has been combined with
the process-independent NLL soft resummation in the initial condition of
the perturbative fragmentation function.

I have presented the $b$-quark energy spectrum in Higgs decay, which exhibits 
a remarkable effect of the implemented collinear and soft resummations.
In particular, soft resummation  smoothens the distribution at large
$x_b$ and  
the dependence of the prediction on factorization and renormalization
scales turns out to be very little.
The dependence of the predicted spectra on the Higgs mass is very small
as well.
I have also compared the $b$-energy spectrum in $H\to b\bar b$ with the
ones yielded by top decay and $e^+e^-\to b\bar b$,
which have been provided with NLL collinear and soft resummations
in the framework of perturbative fragmentation.

I have then considered $b$-flavoured $B$-hadron production in Higgs decay in
both $x_B$ and moment space.
I have fitted a few hadronization models to 
ALEPH and SLD data in $x_B$-space and used the best-fit parameters
to predict the $B$ spectrum in $H$ decays. 
For this procedure to be consistent, the perturbative processes
$e^+e^-\to b\bar b$ and $H\to b\bar b$ have been described using the
same kind of calculation.
The fits have shown that the power law with two tunable parameters and 
the Kartvelishvili model give good combined 
fits of ALEPH and SLD data, while
the Peterson model is not capable of reproducing the data.
Moreover, implementing soft-gluon resummation in the coefficient function
has been essential to be able to describe the data.
The $B$-hadron spectra in Higgs decay according
to the power law and to the Kartvelishvili model
are in statistical agreement.
The comparison of $B$ spectra in $H\to b\bar b$, $t\to bW$ and 
$e^+e^-\to b\bar b$ exhibits similar features to the parton-level one.
I have also made predictions for the first four moments of the $B$ 
spectrum in $H\to b\bar b$ processes, using DELPHI data
on the $B$ moments in $e^+e^-$ annihilation.

In conclusion, the presented calculation allows the performance of precise
predictions for $b$-quark and $B$-hadron production in Higgs decay and could
be applied for analyses of Higgs phenomenology at the Tevatron and
ultimately at the LHC. The computed NLO matrix elements, along
with NLL DGLAP evolution and soft resummation, can be used for
further investigations of other observables in Higgs decay.
In particular, studies of the transverse momentum distributions
of $b$ quarks and $b$-flavoured hadrons are under way.
Moreover, 
as the NNLO initial condition of the perturbative
fragmentation function has recently been computed \cite{alex}, 
the coefficient function may be calculated
to NNLO as well and a resummation of collinear and soft terms to 
next-to-next-to-leading logarithmic accuracy
can be made.

\section*{Acknowledgements}
I am grateful to M. Cacciari, who provided me
with the computer code to perform inverse Mellin transforms and fits
of hadronization models to $e^+e^-$ data.
I thank S. Catani, L. Magnea,
F. Maltoni, S. Moretti and M.H. Seymour for several discussions 
on these and related topics. 
I acknowledge V. Ravindran for pointing out an error in Eq.~(10) in the
first version of the paper.

\end{document}